\documentclass[a4paper,10pt,oneside]{article}
\usepackage[english]{babel}
\usepackage{graphicx}
\usepackage{wrapfig,epsfig,epsf}

\begin{document}

\title{  Dynamical Instability of Laminar Axisymmetric Flow of Perfect Fluid with Stratification. }

\author{ 
Zhuravlev V.V.$.^{1}$  ,  Shakura N.I.$.^{1,2}$
} 

\date{ \it \small
\footnote {e-mail: slava@xray.sai.msu.ru}
1) Sternberg Astronomical Institute, Moscow, Russia 
\\
2) Max Planck Institute for Astrophysics, Garching, Germany 
\\
\vspace{0.2cm}
accepted to Astronomy Letters
}

\maketitle

\bigskip
key words: hydrodynamics -- instability -- non-homoentropic flows.

\bigskip

\large

\begin{abstract}

    The instability of non-homoentropic axisymmetric flow of perfect fluid with respect to non-axisymmetric infinitesimal perturbations 
was investigated by numerical integration of hydrodynamical differential equations in two-dimensional approximation. 
The non-trivial influence of entropy gradient on unstable sound and surface gravity waves was revealed. In 
particular, both decrease and growth of entropy against the direction of effective gravitational acceleration $g_{eff}$ 
give rise to growing surface gravity modes which are stable with the same parameters in the case of homoentropic flow. 
At the same time increment of sound modes either grows monotonically while the rate of entropy decrease against $g_{eff}$ gets higher 
or vanishes at some values of positive and negative entropy gradient in the basic flow. The calculations have showed
also that growing internal gravity modes appear only in the flow unstable to axisymmetric perturbations. At last, the 
analysis of boundary problem with free boundaries uncovered that's incorrect to set the entropy distribution according
to polytropic law with polytropic index different from adiabatic value, since in this case perturbations don't satisfy the 
free boundary conditions.

\end{abstract}

\newpage

\section{Introdution}

This paper is natural extension of the two previous, in which the dynamical instability of axisymmetric flows 
was considered in approximation of incompressible perfect fluid  (Zhuravlev \& Shakura, 2007a - hereafter ZS1) and 
with finite compressibility (Zhuravlev \& Shakura, 2007b - hereafter ZS2). Papaloisou and Pringle (1984, 1985, 1987) were the first
who investigated the instability to non-axisymmetrical perturbations being under discussion in this paper. Their discovery entailed 
numerous explorations, which are mentioned partially in the review by Papaloisou \& Lin (1995). The growth of perturbations 
in axisymmetric flows with free boundaries is an important problem for astrophysics and is of special interest in the view of 
question how angular momentum transports outwards in accretion flows. Though the realistic astrophysical conditions force us to 
include the effects of radiative heating and cooling, magnetic fields, relativity and so forth, it's still necessary to study 
the accretion flow as a shear flow in the simple hydrodynamical approximation. 
In the most of such investigations the angular velocity profile was set by a power law $\Omega \propto r^{-q}$ with $1.5<q\leq 2.0$, which means 
that $g_{eff}$ attains the greatest value at the boundaries. In ZS1 and ZS2 alone with the power law we considered another 
$\Omega$ profile which implies that $g_{eff}=0$ at the boundaries. It was revealed that the absence of effective gravity 
at the edge of the flow essentially affects the growing perturbation modes. In particular, it causes the disappearance of 
unstable sound modes so that the flow with angular velocity profile being close to the Keplerian one 
becomes stable to infinitesimal perturbations. Additionally, the special attention was paid to the influence of vorticity 
gradient on the instability. 
In the present paper we are going to study how stratification of the basic flow affects the growing sound and surface gravity
waves and make an attempt to find unstable internal gravity waves. Following the other authors we consider the barotropic 
basic flow, i.e. the entropy distribution is set in such way that isopycnic surfaces coincide with isobaric ones. It allows 
to regard angular velocity as function of radial direction. Discussing the instability of stratified axisymmetric flows we should
refer, for instance, to the paper by Frank \& Robertson (1988) who considered the instability of tori with random initial small 
perturbations and Kojima et al.  (1989) who studied toroidal as well as cylindrical flows. The latter are two-dimensional analogue of
real flows in point-mass gravitational  potential. We should note here that in both cases the obtained results proved to be similar
and 3-dimensional growing modes of perturbations in the toroidal flow turned out to have no essential dependence on 
vertical direction. The authors explained this by the fact that  Reinolds stress didn't contain vertical perturbation velocity 
for barotropic configurations with $\Omega(r)$. Further, Glatzel (1990) considered the instability of cylindrical and plane-parallel 
flows in approximation of small extent of a shear layer. In order to exclude the growing sound and surface gravitational modes, he
assumed an incompressible fluid and rigid boundaries so that stratification was entered by the density profile. The instability he
found was explained as the result of over-reflection and coupling of internal gravitational modes. Later Ghosh \& Abramowicz (1991)
studied the cylindrical flow consisting of two fluids with different constant densities. The configuration was Rayleigh-Taylor stable.   
Alone with modified branch of growing surface gravity mode being consequence of the free boundaries (Blaes \& Glatzel, 1986)
they found another branch of increments appeared owing to density discontinuity. This instability is caused by growing 
internal gravitational mode, which, however, is fully analogous to the growing surface mode, since the boundary between two 
fluids differs from the free boundary just by the finite ratio of densities. 

We should also mention about the results of Lovelace et al. (1999) and Li et al. (2000) where in the 2-dimensional geometry the instability 
of thin Keplerian disks with the local maximum of entropy was investigated. In particular, the local dispersion relation for non-axisymmetric 
perturbations similar to Rossby waves dispersion relation  was derived. At last, Klahr \& Bodenheimer (2003) studied the instability of 
Keplerian disks with entropy decreasing outwards.

As it's well known, the stability of stratified flows is governed by Richardson criterium. Initially it was formulated for plane-parallel 
flows (see, for instance, Howard, 1961). In this case, the sufficient condition for stability is $Ri>1/4$ in each point, where $Ri$ - 
is the Richardson number. The generalization of the Richardson criterium for baroclinic rotating flows, when angular velocity depends
also on the vertical direction, was found by Fujimoto (1987) for incompressible fluid and by Hanawa (1987) for compressible fluid.

In the present work we follow other authors and consider the simple 2-dimensional geometry, i.e. assume the cylindrical flow and 
perturbations with zero vertical velocity. In the first section we'll write down the basic equation for infinitesimal perturbations, 
formulate the free boundary conditions and set the main features of the basic flow - angular velocity profile and entropy 
distribution. Then we'll touch on a numerical method, in particular, discuss the question when the perturbations 
can satisfy the boundary conditions and finally will present our results.

\section{Equation and boundary conditions.}

\subsection{Derivation of main equation}

An isentropic motion of fluid is governed by the system of equations:
(see, for example, Brekhovskikh \& Goncharov, 1982):

$$
\frac{\partial {\bf v}}{\partial t} + ({\bf v} \nabla) {\bf v} = - \frac{\nabla p}{\rho} - \nabla \Phi , 
$$
\begin{equation}
\label{PertFlow2}
\frac{\partial \rho}{\partial t} + \nabla {(\rho \bf v )} = 0,
\end{equation}
$$
\frac{d}{dt}\left( \frac{p}{\rho^\gamma} \right)=0,
$$
where $\Phi$ is an external gravitational potential; 
The last equation represents the conservation of entropy in the particle of fluid with the perfect gas equation of state -
$p=K \, e^{S/c_v} \rho^\gamma$,
with $\gamma=c_p/c_v$ - adiabatic index, $S$ и $c_v$ - specific entropy and heat capacity for constant volume.

The linearization procedure gives the equations for small perturbations:
$$
\frac{\partial \delta {\bf v}}{\partial t} + ({\bf v} \nabla) \delta {\bf v} + 
({\delta \bf v} \nabla){\bf v}  =
-\frac{\nabla \delta p}{\rho} + \frac{\delta \rho}{\rho}\frac{\nabla p}{\rho},  
$$
\begin{equation}
\label{EqPerts2}
\frac{\partial \delta \rho}{\partial t} + \nabla (\rho \delta {\bf v}) + \nabla ( \delta \rho {\bf v}  ) = 0
\end{equation}
$$
\frac{\partial }{\partial t} \left ( \frac{\delta p}{\rho^\gamma} - \gamma \frac{\delta\rho}{\rho} \frac{p}{\rho^\gamma} \right ) +
({\bf v} \nabla) \left ( \frac{\delta p}{\rho^\gamma} - \gamma \frac{\delta\rho}{\rho} \frac{p}{\rho^\gamma} \right ) + 
(\delta{\bf v} \nabla)  \frac{p}{\rho^\gamma}=0,
$$
where the third one implies the relation between perturbations of density and pressure.

\vspace{0.5 cm}

In our approach all quantities are functions of $r$, so for Eulerian perturbations with $\delta v_z=0$ in the cylindrical coordinates 
the system (\ref{EqPerts2})  will take the form:

$$
\frac{\partial \delta v_r}{\partial t} + 
\Omega \frac{\partial \delta v_r}{\partial \varphi} - 
2\Omega \delta v_\varphi =
-\frac{1}{\rho}\left( \frac{\partial\delta p}{\partial r} - \frac{\delta \rho}{\rho}\frac{dp}{dr} \right ), 
$$
\begin{equation}
\label{secEq2}
\frac{\partial \delta v_\varphi}{\partial t} +
\Omega \frac{\partial \delta v_\varphi}{\partial \varphi} + 
\frac{1}{r}\frac{d}{dr}  ( \Omega r^2  ) \delta v_r =
-\frac{1}{r\rho}\frac{\partial \delta p}{\partial \varphi},     
\end{equation}
$$
\frac{\partial \delta\rho}{\partial t} + \Omega \frac{\partial \delta\rho}{\partial \varphi} +  
\frac{1}{r} \frac{\partial}{\partial r} \left ( r\rho\delta v_r \right ) + 
\frac{1}{r}\frac{\partial}{\partial \varphi} \left (\rho \delta v_\varphi \right ) = 0,
$$
$$
\frac{\partial}{\partial t} \left ( \delta p - \gamma p \frac{\delta\rho}{\rho} \right ) +
\Omega \frac{\partial }{\partial \varphi} \left ( \delta p - \gamma p \frac{\delta\rho}{\rho} \right ) + 
\rho^\gamma \delta v_r \frac{\partial}{\partial r} \left ( \frac{p}{\rho^\gamma} \right ) = 0
$$

\noindent
We'll be looking for a solution in the form of normal modes:

\begin{equation}
\label{exp2}
\begin{array}{l}
{\delta \bf v} = {\bf \bar v}(r) e^{-i(\omega t - m \varphi)}, \\
\delta p = \bar p (r) e^{-i(\omega t - m \varphi)} \\
\delta \rho = \bar \rho (r) e^{-i(\omega t - m \varphi)} 
\end{array}
\end{equation}
here $m$ - is the azimuthal wavenumber, $\omega=\omega_r + i\, \omega_i$ - complex frequency, 
which for growing modes has the positive imaginary part, i.e. increment of mode. 
Possible $\omega$ has to be determined from the solution of the boundary problem coming to the integration of 
ordinary differential equation with the boundary conditions in the boundary points $r_1$ and $r_2$ for functions
${\bf \bar v}$, $\bar p$ и $\bar \rho$, which, in general, are complex. 

\noindent
The substitution (\ref{exp2}) into (\ref{secEq2}) yields:
$$
(-i\omega + im\Omega)\bar v_r - 2\Omega \bar v_\varphi = 
-\frac{1}{\rho}\left( \frac{d \bar p}{d r} - \frac{\bar \rho}{\rho}\frac{dp}{dr} \right ),
$$
\begin{equation}
\label{thirdEq2}
(-i\omega + im\Omega)\bar v_\varphi +  
\frac{1}{r}\frac{d}{dr}\left(\Omega r^2\right) \bar v_r = 
- \frac{im}{r\rho} \bar p
\end{equation}
$$
(-i\omega + im\Omega)\bar\rho +  \frac{1}{r}\frac{d}{dr}(r\rho\bar v_r) + \frac{im\rho}{r}\bar v_\varphi = 0,
$$
$$
(-i\omega + im\Omega) \left ( \bar p - \gamma p \frac{\bar\rho}{\rho} \right ) +
\rho^\gamma \bar v_r \frac{d}{d r} \left ( \frac{p}{\rho^\gamma} \right ) = 0
$$

From the last expression in (\ref{thirdEq2}) we'll get the relationship between $\bar p$ and $\bar \rho$:
\begin{equation}
\label{dp_drho}
\frac{\bar \rho}{\rho} = \frac{1}{a^2}\frac{\bar p}{\rho} - \bar\xi_r\, \frac{N^2}{g_{eff}},
\end{equation}
where $a^2=\gamma p/\rho$ - is the squared sound velocity, 
\begin{equation}
\label{g_eff}
g_{eff} = (\Omega^2 - \frac{GM}{r^3})\,r =  \frac{1}{\rho}\frac{d p}{d r} 
\end{equation}
- the effective gravitational acceleration and
 $\bar\xi_r=i\bar v_r / (\omega-m\Omega)$ gives the radial dependence of radial component of the Lagrangian displacement : $\xi_r( \bf{r}, \, t)  = \bar \xi_r(r)\, e^{-i(\omega t-m \varphi)}$.
At last,
$$
N^2 = g_{eff}\, \left ( \frac{1}{\rho}\frac{d\rho}{dr} - \frac{g_{eff}}{a^2} \right )
$$
- is the squared buoyancy, or Brunt-Vaasala, frequency which is the characteristic internal oscillation frequency in 
medium at rest due to the entropy gradient and gravitational force. Clearly, for constant entropy $dp=a^2d\rho$, $N^2=0$, and the relation between the Eulerian perturbations of density and 
pressure comes to the expression one has for homoenropic configuration, when the perturbed flow as well as the unperturbed one are 
barotropic: $\bar p =a^2 \bar\rho$.

Then, the equation of state  $p=Ke^{(\gamma-1)\,s}\rho^{\gamma}$, where $s$ -
entropy in the units of the universal gas constant $\Re$ and squared sound velocity $a^2 = \gamma p /\rho$, 
gives the following differential relations:
\begin{equation}
\label{diff_relations}
\frac{da^2}{dr} = (\gamma-1) \, g_{eff} + a^2\,\frac{\gamma-1}{\gamma}\,\frac{ds}{dr}, 
\end{equation}
$$
\frac{1}{\rho}\frac{d\rho}{dr} = \frac{g_{eff}}{a^2} - \frac{\gamma-1}{\gamma}\frac{ds}{dr}
$$
(\ref{diff_relations}) shows that squared buoyancy frequency is simply

\begin{equation}
\label{BruntVaisala}
N^2 = -\frac{\gamma-1}{\gamma} \, g_{eff} \, \frac{ds}{dr},
\end{equation}

So for stratified medium the eq.  (\ref{dp_drho}) is modified just by the additional 
term proportional to the entropy gradient. 
\noindent
Finally, from the system (\ref{thirdEq2}) we obtain the equation for infinitesimal perturbations with respect to $\bar p / \rho$:
$$
\frac{d}{dr} \left ( \frac{r\rho}{D} \, \frac{d}{dr}\left( \frac{\bar p}{\rho}  \right )  \right ) - 
\left [   \frac{2m}{\omega - m\Omega}\, \left (  \frac{d}{dr} \left ( \frac{\Omega \rho}{D} \right ) - 
\frac{2\Omega\rho}{D}\,\frac{N^2}{g_{eff}} \right ) \right . + 
$$
\begin{equation}
\label{EntalpyEq2}
\frac{m^2}{D r}\, \rho \left ( 1 - \frac{N^2}{(\omega-m\Omega)^2}  \right )   
\left . + r\rho \left ( \frac{1}{a^2} + \frac{1}{D}\,\left ( \frac{N^2}{g_{eff}}\right )^2\,  \right ) -
\frac{d}{dr}\left ( \frac{r\rho}{D}\,\frac{N^2}{g_{eff}}  \right )  \right ] \, \frac{\bar p}{\rho} = 0,
\end{equation}

\vspace{0.5cm}

where $D = D_h + N^2$, $D_h=\kappa^2 - (\omega-m\Omega)^2$ - the quantity signed as ''D'' for homoentropic flow in ZS2.
The squared epicyclic frequency is the same as before:
$$
\kappa^2 = \frac{2\Omega}{r} \frac{d}{dr} \left ( \Omega r^2 \right ).
$$

\noindent
Excluding the difference in notations, the main equation (\ref{EntalpyEq2}) coincides with one used by Lovelace et al. (1999)
for perturbations integrated in the z-direction. 
For $N^2=0$ the eq. (\ref{EntalpyEq2}) comes to the equation for the Eulerian perturbation of enthalpy in homoentropic fluid used in ZS2. 
In the limit $\gamma\to\infty$  the equation (\ref{EntalpyEq2}) determines the behavior of small perturbations in stratified incompressible flow. 
At the same time the entropy gradient is connected directly with the variable density: 
\begin{equation}
\label{entr_incompr}
\frac{ds}{dr} = - \frac{1}{\rho} \, \frac{d\rho}{dr},
\end{equation}
and the squared buoyancy frequency is:
\begin{equation}
\label{BV_incompr}
N^2 = g_{eff} \, \frac{1}{\rho} \, \frac{d\rho}{dr},
\end{equation}

\vspace{1cm}
\noindent
At last, we write here the expressions for $\bar v_r$ and $\bar v_\varphi$, in terms of $\bar p / \rho$:
$$
\bar v_r = \frac{i}{D} \left [ (\omega-m\Omega)\, \frac{d}{dr}\left ( \frac{\bar p}{\rho} \right ) - \left ( \frac{2m\Omega}{r}\, 
- (\omega-m\Omega)\,\frac{N^2}{g_{eff}} \right ) \left ( \frac{\bar p}{\rho} \right ) \right ],
$$

$$
\bar v_\varphi = \frac{1}{D} \left [ \frac{\kappa^2}{2\Omega}\,\left ( \frac{d}{dr}\left ( \frac{\bar p}{\rho} \right ) +
\frac{N^2}{g_{eff}} \left ( \frac{\bar p}{\rho} \right ) \right ) - \frac{m}{r} \, 
\left ( \omega-m\Omega - \frac{N^2}{\omega-m\Omega} \right ) \, \left ( \frac{\bar p}{\rho} \right )  \right ]
$$

\subsection{The boundary conditions}

We have to impose boundary condition for $\bar p /\rho$.
At the free boundary for perturbed as well as for unperturbed flow $h=a^2=p=0$. Here $h$ - is an enthalpy of perfect gas.
This implies vanishing of the Lagrangian perturbation of enthalpy at the boundary of the unperturbed flow $\Delta h|_{b} = 0 $. 
Further, for the fixed fluid particle $dh=dp/\rho + TdS=dp/\rho$ since the motion is isentropic. 
Consequently, $\Delta h = \Delta p / \rho=0$ at the free boundary.
Using the relation between small Eulerian and Lagrangian quantities (look Tassoul, 1978) we have: 
\begin{equation}
\label{Boundary2}
\frac{\delta p}{\rho} + \xi_r\frac{1}{\rho}\frac{dp}{dr}\left |_{гр} \right . = 0\,
\end{equation}

The eq. (\ref{Boundary2}) yields the boundary condition for $\bar p /\rho$.
Here it's necessary to use the relation between $\bar \xi_r$ and $\bar v_r$ have been used to obtain (\ref{dp_drho}) and expression for $\bar v_r$ 
which was written at the end of the last section:

\begin{equation}
\label{Boundary22}
g_{eff}\,\frac{d}{dr}\left ( \frac{\bar p}{\rho}  \right ) - 
\left ( D_h + \frac{g_{eff}}{r}\,\frac{2m\Omega}{\omega-m\Omega} \right )\, \left ( \frac{\bar p}{\rho}  \right ) \quad
\Biggl{|_{r_1, r_2}} \Biggr .= 0
\end{equation}
One can check that (\ref{Boundary22}) is exactly the same as the boundary condition for 
the homoentropic flow used in ZS2. In particular,  the eq. (\ref{Boundary22}) doesn't contain terms with variable entropy.

\vspace{1cm}

\section{Angular velocity and entropy profiles}

The rotation curve is determined by gravitation and the pressure gradient. In the previous research (ZS1 and ZS2) we used 
the two profiles of angular velocity of the basic flow:

\begin{equation}
\label{rotation11}
\Omega(r) = \Omega_0\,\, \left ( \left ( \frac{r}{r_0}  \right )^{-3} +
\frac{K}{r/r_0} \sin \left (
2\pi\frac{r-r_1}{r_2-r_1} \right )
\right )^{1/2}
\end{equation}
\qquad\qquad\qquad\qquad\qquad  
\begin{equation}
\label{rotation21}
\Omega(r) = \Omega_0 \left ( \frac{r}{r_0}  \right )^{-q} 
\end{equation}
here $r_1$ and $r_2$ - are the boundary points of the flow, $r_0$ - point where the matter rotates with the Keplerian frequency $\Omega_0$. 
The coefficient $K$ sets the deviation of $\Omega(r)$ from the Keplerian profile. 
All frequencies will be in units of $\Omega_0$ and all the distances - will be given in units of $r_0$ below. 
To define the radial extent of the flow we'll use the quantity
$w=(r_2-r_1)/r_0 \times 100\%$.

The power law (\ref{rotation21}) has been traditionally used in the research of axisymmetric flow instabilities.
This profile is characterized by the maximum of $g_{eff}$ at the edge of the flow for
$q>3/2$. The rotation according to (\ref{rotation11}) implies that   
$g_{eff}$ obeys the sinusoidal law and vanishes at the boundaries. 
In ZS2 it was showed that $g_{eff} \to 0$ in $r_1$ and $r_2$ entails the absence of growing sound modes. 
For consistency we take here the same rotation profiles  
although the main attention will be paid to the power law. 
 
To solve the equation (\ref{EntalpyEq2}) one should calculate first $a^2(r)$. 
Integrating the first expression in (\ref{diff_relations}) we have: 
\begin{equation}
\label{sound_vel}
a^2 = e^{(\gamma-1)\,s/\gamma} \, 
\left [ \int (\gamma-1)\, g_{eff} \,\, e^{-\,(\gamma-1)\,s/\gamma}\, \, dr \, + C_1  \right ], 
\end{equation}
where $g_{eff}$ is defined by the expr. (\ref{g_eff}), and  
$C_1$ for specific $w$ is determined from the condition that $a^2$ must vanish at $r_1$ and $r_2$. 
Let's notice that for profile(\ref{rotation11}) the boundary points are defined beforehand, since $r_0=(r_2+r_1)/2$. 
For (\ref{rotation21})  $r_1$ and $r_2$ must be calculated together with the constant $C_1$.

For the certain parameters $a^2(r)$ was determined by the Chebyshev polynomials approximation of the tabular function obtained by the numerical integration 
of the equation (\ref{sound_vel}).
Note that we could avoid difficulties with $a^2$ calculation using an appropriate angular velocity profile. But in this case 
we would miss the possibility to compare the new results with the previous ones found in SZ1 and ZS2 in the limit of uniform entropy distribution. 

Besides, we'd lose a chance to explore the effects induced by stratification purely, i. e. with the same distribution of effective gravity and 
vorticity in the basic flow.

In the limit of $\gamma \to \infty$ one doesn't have to find explicit dependencies of the flow quantities (such as $p$ or $\rho$) on $r$ to solve (\ref{EntalpyEq2}).
However, in case of power law rotation profile we need to determine the boundary points, what can be done using the stationarity condition.  
Since  $g_{eff}=\frac{1}{\rho}\frac{dp}{dr}$, we have:
\begin{equation}
\label{p_incompr}
p = \int g_{eff}\,\rho\, dr \,+\, C_2
\end{equation}
To integrate the equation (\ref{p_incompr}), one should know $\rho(r)$. For the incompressible stratified medium we get $\rho(r)$ by integration of the equation (\ref{entr_incompr}) 
with the certain entropy profile.

\paragraph{Entropy profile\\}

The entropy distribution was defined as follows:
\begin{equation}
\label{entropy}
s(r) = s_0 - s_1 \, (r/r_0-1)^2,
\end{equation}
where as before, $r_0$ - is the point in the flow with Keplerian rotation, so that $g_{eff}=0$, 
$s_0,\,s_1$ - are the constants. This form is convenient because when $s_1 \ne 0$ the entropy gradient is directed in the same way with respect to 
${\bf g}_{eff}$ everywhere in $(r_1,\,r_2)$, in the other words $N^2$ has the same  sign going to the zero in the vicinity of $r_0$. 
Moreover,  the function (\ref{entropy}) is monotonous on the each side of $r_0$, what allows to set entropy before calculation of the boundary points necessary for
the power law rotation profile. 
Note that for the fixed $w$ 
the result of calculations doesn't depend on $s_0$. 
Indeed, adjusted for the dep.  (\ref{entropy})  squared sound speed (\ref{sound_vel}) is:
$$
a^2(r) = F_1(r) + L(s_0,C_1)\,F_2(r)
$$
Now, if in the certain boundary point $a^2=0$, we find $L(s_0,C_1)$. It means, that 
$s_0$ controls only $C_1$, and leaves $a^2$ to be unchanged.
Then, since  the equation (\ref{EntalpyEq2}) contains only the derivatives of $s(r)$, we concluded that eigenvalues of the boundary problem 
are not affected by the value of  $s_0$. So $s_1$ is the only parameter that characterizes the stratification of the flow.

\vspace{0.5cm}
Opposite to non-axisymmetric  an axisymmetric perturbations obey the Heiland's criterium (Tassoul, 1978).
Namely, for considered stationary flow the necessary and sufficient condition of it's stability to perturbations with axial symmetry is the inequality: 
\begin{equation}
\label{heiland}
\kappa^2+N^2>0
\end{equation} 
Thus the stratification with $s_1<0$ contributes to the stabilization of the flow (to axisymmetric perturbations), and vice versa,  
with $s_1>0$ - to the instability in comparison with the homoentropic case.  
For  $\Omega(r)\propto r^{-q}$ with  $q<2$ and $s_1>0$ the flow with sufficiently small  $w$ will always satisfy to Heiland's criterium. 
However, the increase of $w$ will cause the break of (\ref{heiland}) close to the boundaries, 
since the entropy gradient goes up while moving away from $r_0$. 

One more criterium is valid for stratified flows. It is the Richardson criterium.
We've mentioned the authors who generalized it for baroclinic axisymmetric flows. For barotropic configuration we consider here it yields that 
\begin{equation}
\label{richardson}
Ri=\frac{N^2}{\left ( r d\Omega/d r\right )^2} > \frac{1}{4}
\end{equation} 
everywhere in $(r_1,\,r_2)$ is the sufficient condition of the stability 
(see also Glatzel, 1990).  $Ri$ - is the Richardson number. It's clear that (\ref{richardson}) is never fulfilled here, since
for any $s_1$ $Ri\to 0$ while $r \to r_0$.

\vspace{1cm}

\section{Numerical calculation}
We'll solve the boundary problem consisting of the equation
(\ref{EntalpyEq2}) and boundary conditions 
(\ref{Boundary22}) at the interval $(r_1,\,r_2)$ for compressible and incompressible fluid.
 Let's rewrite (\ref{EntalpyEq2}) in the form convenient for integration:
$$
\left ( \frac{\bar p}{\rho}\right )^{\prime\prime} + \left ( \frac{1}{r} - \frac{D^\prime}{D} + \frac{N^2}{g_{eff}} + \frac{g_{eff}}{a^2} \right )
\left ( \frac{\bar p}{\rho} \right )^\prime  - 
$$
\begin{equation}
\label{num_eq}
\left [ \frac{2m}{\omega-m\Omega} \left ( \frac{\Omega^\prime}{r} + \frac{\Omega}{r} \left ( \frac{g_{eff}}{a^2} -
\frac{D^\prime}{D} -\frac{N^2}{g_{eff}} \right )  \right )  + \frac{m^2}{r^2} \left ( 1 - \frac{N^2}{(\omega-m\Omega)^2} \right ) + \right .
\end{equation}
$$
\left . \frac{D}{a^2} - \left ( \frac{N^2}{g_{eff}} \right )^\prime - \frac{N^2}{g_{eff}} \left ( \frac{1}{r} - 
\frac{D^\prime}{D} + \frac{g_{eff}}{a^2} \right ) \right ] \left (
\frac{\bar p}{\rho} \right ) = 0
$$
with the finite compressibility and 
$$
\left ( \frac{\bar p}{\rho}\right )^{\prime\prime} + \left ( \frac{1}{r} - \frac{D^\prime}{D} + \frac{N^2}{g_{eff}} \right )
\left ( \frac{\bar p}{\rho} \right )^\prime  - 
\left [ \frac{2m}{\omega-m\Omega} \left ( \frac{\Omega^\prime}{r} - \frac{\Omega}{r} \left ( 
\frac{D^\prime}{D} + \frac{N^2}{g_{eff}} \right )  \right ) + \right .
$$
\begin{equation}
\label{num_eq_incompr}
\left . \frac{m^2}{r^2} \left ( 1 - \frac{N^2}{(\omega-m\Omega)^2} \right ) 
 - \left ( \frac{N^2}{g_{eff}} \right )^\prime - \frac{N^2}{g_{eff}} \left ( \frac{1}{r} - 
\frac{D^\prime}{D} \right ) \right ] \left ( \frac{\bar p}{\rho} \right ) = 0
\end{equation}
in the incompressible limit. In the equations (\ref{num_eq}) and (\ref{num_eq_incompr}) the derivative on $r$ is marked by a stroke. 
Since we're interested in the growing modes the integration can be implemented along the real axis according to the Lin's rule (1945).
Each of the equations (\ref{num_eq}) and (\ref{num_eq_incompr}) have been separated into the real and imaginary parts, so that  
the system of four first order equations consisting of the real quantities have been integrated.

For $\omega_i \ne 0$ the equation (\ref{num_eq_incompr}) doesn't contain singular points at real axis so the boundary problem 
have been solved as in ZS1. The four specified linearly independent vectors have been used as initial conditions in $r_1$ 
to obtain the corresponding solutions in $r_2$. The boundary conditions in $r_1$ and $r_2$ give then four algebraic equations
and the eigenvalues should be obtained by setting the determinant of corresponding matrix equal to zero.
The finite compressibility involves the sound speed to vanish at the boundaries. The main equation contains now 
singularities in $r_1$ and $r_2$ (for power law rotation it's the first order poles and for (\ref{rotation11}) profile
it's the second order poles) so the numerical algorithm have to be adjusted appropriately.  

As in ZS2 the solution should be expanded to the generalized series in the vicinity of $r_1$ and $r_2$:

\begin{equation}
\label{series}
\frac{\bar p}{\rho}  = (r-r_{1,2})^\mu \, \sum_{i=0}^\infty\, a_i (r-r_{1,2})^i
\end{equation}

Substituting 
(\ref{series}) into 
(\ref{num_eq}) and multiplying it by $(r-r_{1,2})^2$ one gets the recurrent expressions for 
$a_i$ and the squared equation for $\mu$ (see ZS2). The last one defines two independent solutions of (\ref{num_eq}) 
close to the boundaries with some $\mu_1$ and $\mu_2$. 
The regularity condition assigns what solution should be chosen.  
After that the numerical integration from left and right boundary gives two solutions in some point inside $(r_1,\,r_2)$.
The coupling condition can be rewritten as the determinant of the certain matrix equal to the zero as above and again gives
an eigenvalues.
At that $\mu_1$ and $\mu_2$ turn out to be independent on the entropy distribution in spite of the fact that 
the recurrent expressions for $a_i$ differ from that in ZS2. For the power law rotation 
$\mu=0$ as before corresponds to the regular solution and
the relation between $a_0$ (value of $\bar p / \rho$ at the boundary) and 
$a_1$ (value of $(\bar p / \rho)^\prime$ at the boundary) as before coincides with the boundary condition (\ref{Boundary22}) 
(As it's shown in ZS2 for the uniform entropy distribution). 

For rotation with quasi-sine deviation from the Keplerian law we choose again the solution with 
$Re(\mu)>1$, what simultaneously implies that it satisfies to the boundary condition when 
$g_{eff}|_{r_{1,2}}=0$. 

\vspace{0.2cm}

\noindent

It's appropriate to mention here one popular way when the stratification of basic flow is set by the relation 
$p \propto \rho^\Gamma$, where $\Gamma \ne \gamma$ 
(see, for instance, the paper by Kojima, 1989). This approach can't be implemented here since at the boundaries
$r \to r_{1,2}$, $\rho \to 0$,  and entropy and its derivatives are not finite in $r_{1,2}$. 
So the coefficients of (\ref{num_eq}) have the singularities also because of the terms that contain entropy gradients.
Then it's easy to demonstrate that, at least for the power law rotation, the relation between $a_0$ and $a_1$ with $\mu=0$ 
won't be equivalent for (\ref{Boundary22}), so that the boundary problem doesn't have the solution. The boundary condition 
is equivalent to the regularity condition only if $s(r)$ with it's derivatives is finite at the boundary.

\section{The results}

\subsection{Instability of flow with the power law rotation profile.}

There are growing sound and surface gravity modes in axisymmetric flow with free boundaries (Papaloisou \& Pringle (1984, 1985, 1987), Blaes \& Glatzel (1986), 
Goldreigh et al. (1986), Glatzel (1987a,b), Sekyia \& Miyama (1988), Kojima (1989) and others). An entropy distribution is supposed to modify mentioned branches 
of instability and possibly the growing internal gravity modes should appear (Glatzel, 1990, 1991). The results of calculations will be presented here mostly 
at the plots, where the curves of increments will be drawn. We also won't limit ourselves with the region stable to axisymmetric perturbations according to 
the Heiland's criterium. In the curves of  $\omega_i(s_1)$ the transition from one region to another will be marked by a stroke.

\begin{figure}
\epsfxsize=12cm
\centerline{\epsfbox{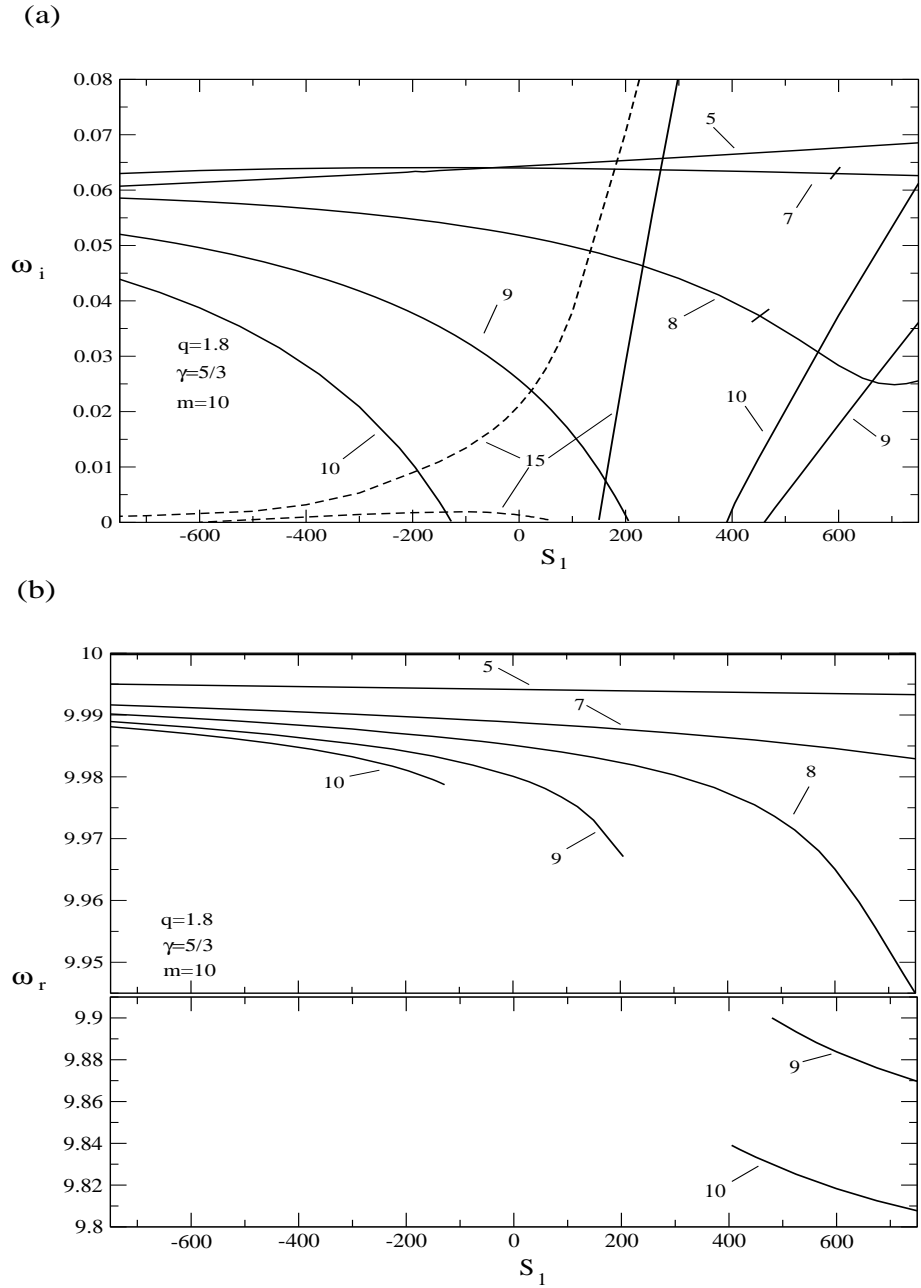}}
\caption
{Increment (pic. (a)) and pattern speed (pic. (b)) of non-axisymmetric modes of perturbations as a function of 
entropy gradient for the power law rotation profile.
The numbers are the values of $w$ in percents.
The solid lines denote the growing surface gravity modes, the dashed lines - the sound modes. The strokes in the curves mark the 
edge of stability to axisymmetric perturbations. For curves corresponding to  
$w=9\%, 10\%, 15\%$, with $s_1>0$ the flow is unstable to axisymmetric perturbations. 
 }
\end{figure}

\paragraph{Calculations with finite compressibility }
In fig. 1 $\omega_i(s_1)$ and $\omega_r(s_1)$ are displayed for the growing surface gravity mode with 
$m=10$, which form so called principal or incompressible branch of instability, since 
$\omega_i>0$ exist also in the limit $\gamma \to \infty$. In the absence of stratification for $q<2.0$ and $\gamma<\infty$
an increment of principal branch disappears at some maximum value of $w$. 
As we can see for small values of $w$ an increment monotonously increases while $s_1$ gets higher, i.e. while the configuration becomes less stable 
according to the Heiland's criterium (\ref{heiland}). However, for greater $w$ minimum appears in the increment curve and then the growing 
incompressible mode disappears at certain $s_1>0$. 
For even greater $w$ surface gravity mode stops growing in the flow with $s=const$, but at the same time increment emerges both for 
the positive ($s_1<0$), and the negative ($s_1>0$) entropy gradient. The last result is consistent with the calculations of Kojima et al. (1989), who made the conclusion that 
entropy growth in the opposite direction to effective gravity (here it's means $s_1<0$) contributes to widening of instability range in the values of $w$.
The similar effect gives the calculation of $\omega_i(s_1)$ for $m=1$ and different values of $q$
(look at fig. 2). One can see that while the vorticity gradient is increasing (i.e. $q$ is decreasing) stabilization sets first in the homoentropic case. Beside that, in fig.2 dashed 
lines denote the additional branches of instability. As it'll be clear below, these are growing internal gravity modes caused by stratification.
However, these modes were found in the present work only in the region unstable with respect to axisymmetric perturbations.
\begin{figure}
\epsfxsize=12cm
\centerline{\epsfbox{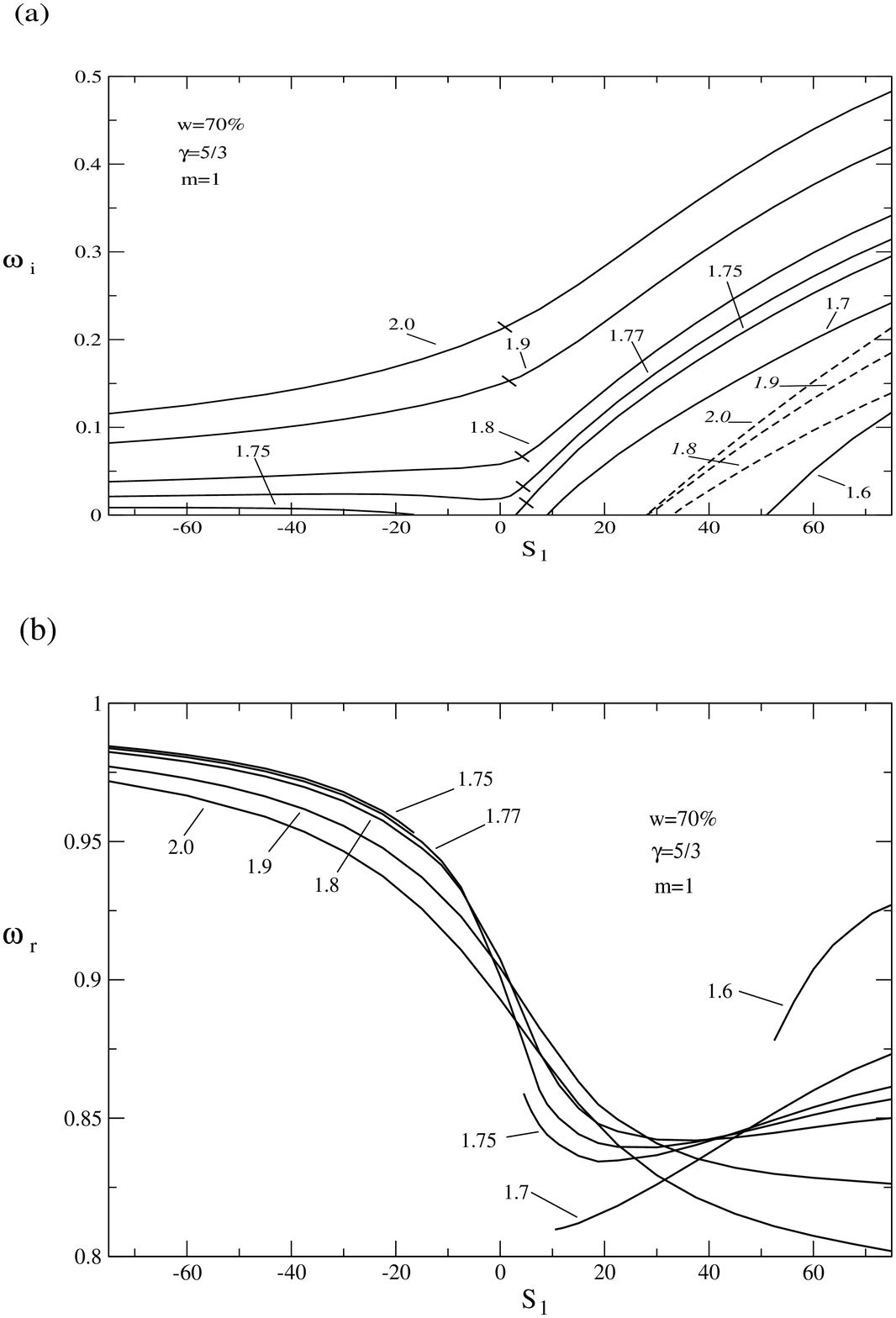}}
\caption
{
The same as in fig. 1, but for fixed $w$ and the dashed lines denote now the growing internal gravity modes.
The numbers are the values of $q$. For curves with  $q=1.6, 1.7$ the flow in unstable to axisymmetric perturbations. 
The curves of pattern speed of internal modes are not depicted because $\omega_r>1$.
}
\end{figure}

In a more intricated manner stratification affects the growing sound modes. 
All the branches of sound modes increments for specific $w$ are depicted in fig. 3. We've used here the results of ZS2, where the detailed calculation of sonic instability 
was implemented, exactly, the calculation of 
$\omega_i(w)$, with $m=10$ and $q=1.58$. We've used the value of $w$ corresponding to the modes coupling for one of the increments. In the
homoentropic flow the modes coupling entails the strong increase of  $\omega_i$ in a narrow range of $w$. In fig. 3 the modes coupling with $s_1=0$ occurs for branch (5).
To the right side from the vertical dashed line the flow is unstable according to the Heiland's criterium. The increment branches fall into three categories:
(1),(8),(2),(7),(3) - the perturbations growth  disappears both at some $s_1<0$ and at some $s_1>0$; 
(4),(6) - increment gradually increases with $s_1$; (5) - corresponds to the modes coupling and $\omega_i(s_1)$ 
behaves in the most tangled way. Exactly, at some value of $s_1=s_1^{cr}>0$ increment (5) vanishes, but comes back at once and grows then with $s_1$. 
The same features of modes coupling arise in dependence $\omega_i(w)$ for homoentropic flow presented in ZS2.  
To check if the presence of  $s=s_1^{cr}$, where $\omega_i\to 0$ is the common peculiarity for modes coupling, 
we've calculated few $\omega_i(s_1)$ for different $w$. 
In can be seen that  all determined $\omega_i(s_1)$, have a break point in increment curve (fig. 4).   

\begin{figure}
\epsfxsize=12cm
\centerline{\epsfbox{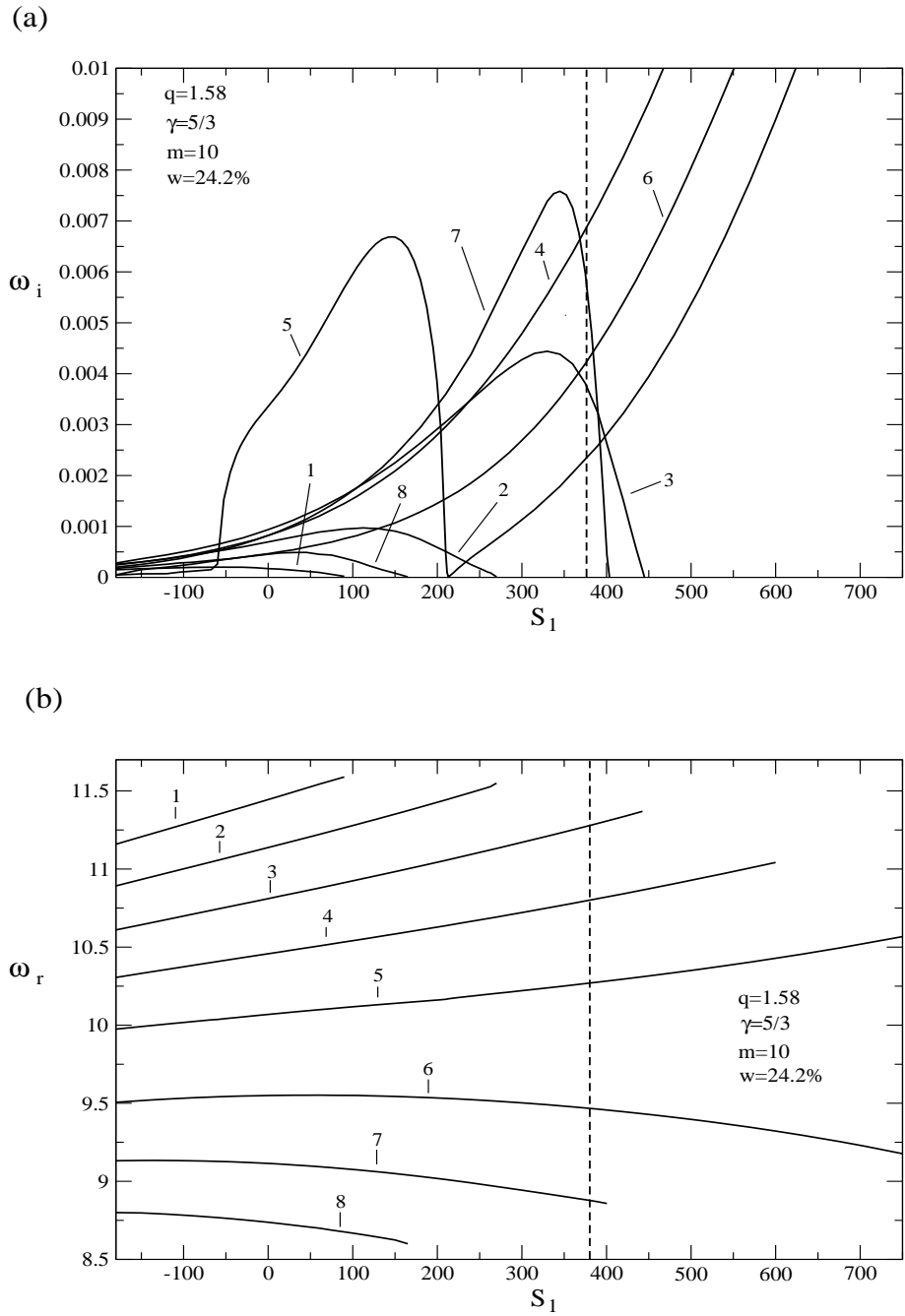}}
\caption
{The same as in the fig. 1, but for fixed $w$ and the increment and pattern speed of sound modes only are presented.
The numerals are the numbers of instability branches. The vertical line marks the limit of stability to axisymmetric perturbations. 
}
\end{figure}

\begin{figure}
\epsfxsize=12cm
\centerline{\epsfbox{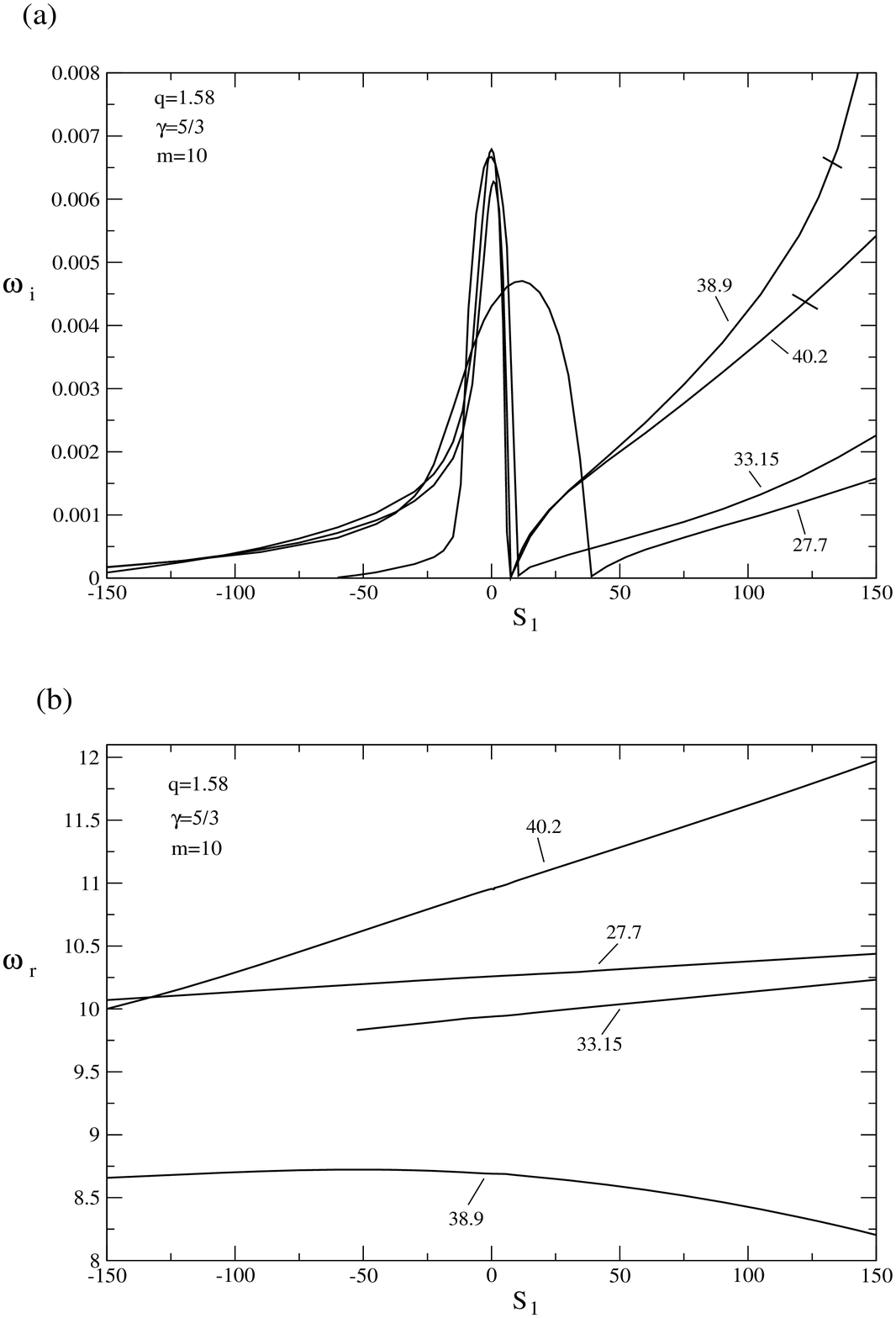}}
\caption
{The same as in the fig. 3 , but for different $w$, signed each curve. The branches presented are the result of 
sound modes coupling (for $s_1=0$). The strokes in the curves denote the edge of stability to 
axisymmetric perturbations. For $w=33.15\%, 27.7\%$ the flow is stable to axisymmetric perturbations in the whole range
of $s_1$.  
}
\end{figure}

At last, branches of sonic instability for 
$m=1$ one can see in fig. 5. The calculations here were mostly carried out in area unstable to axisymmetric perturbations. 
It was revealed that both at some $s_1<0$ and at some $s_1>0$ increment vanishes. In ZS2 we discussed that the growing sound modes with $m=1$ 
arise as a result of two mechanisms: resonant  amplification by the basic flow in critical layer (mechanism Landau) and coupling with decaying surface 
gravity mode, what sets it apart from the above instability being the consequence of coupling of two sound modes. Clearly, the increment behavior 
differs from what we've seen in fig. 4. The main is that in this case for none of the branches increment arises again after vanishing at some $s_1>s_1^{cr}$.

\begin{figure}
\epsfxsize=12cm
\centerline{\epsfbox{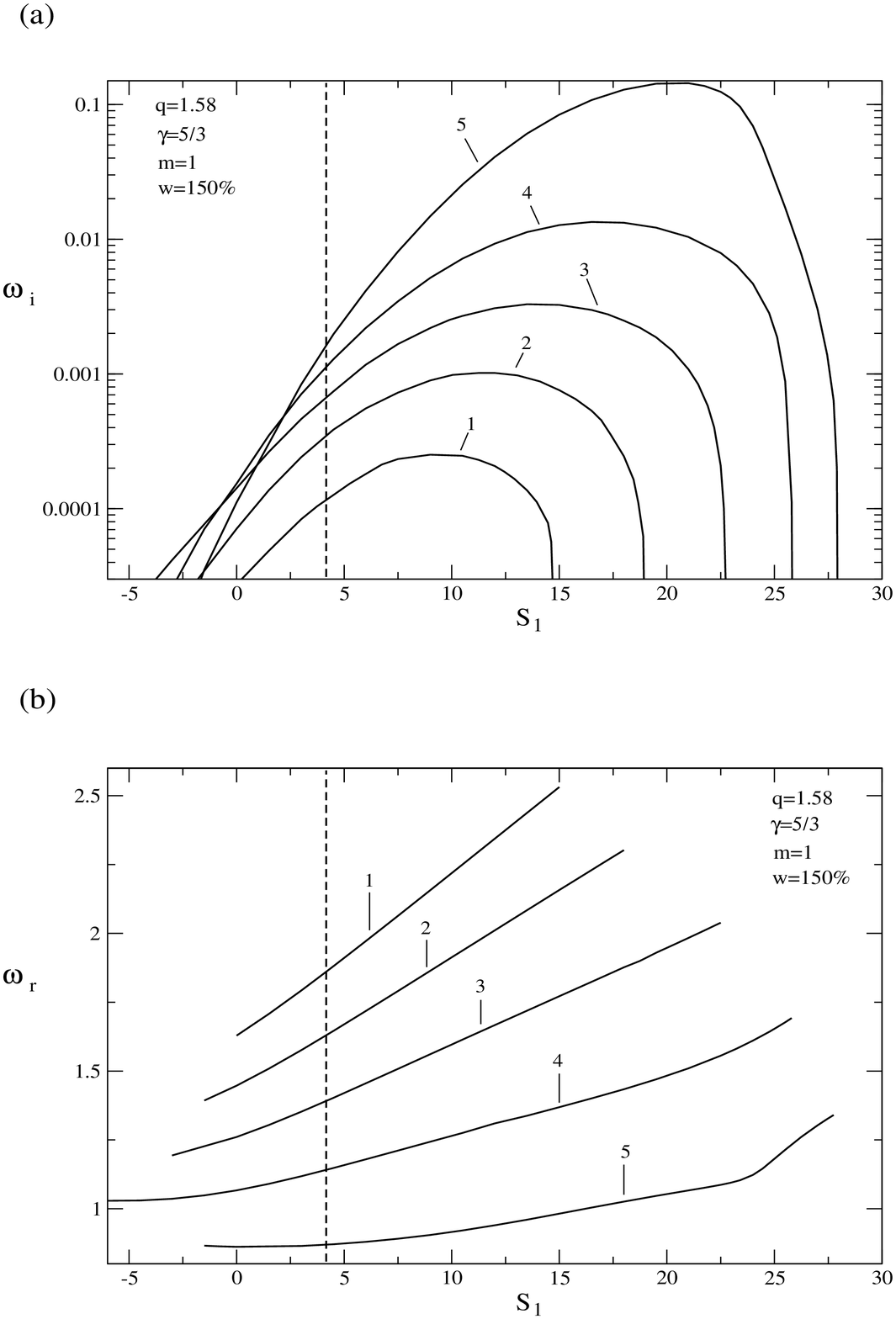}}
\caption
{The same as in the fig. 3 but for $m=1$.
}
\end{figure}

While calculating the presented curves we tried to find the growing internal gravity modes in area stable to axisymmetric perturbations. 
However we failed to do that and the only one instability that arises due to stratification is denoted by dashed lines in fig. 2.      
We've made an attempt then to find growing internal gravity modes in the incompressible limit, since this branch of instability must be insensitive to the compressibility.

\paragraph{Calculations in incompressible limit}
An incompressible limit was used here first of all as an additional way to check the results presented so far, since the numerical 
algorithm differs considerably in that case (look above). In fig.6 we put increment curves with dependence on 
$\gamma$ (pics a \& b) and with dependence on $s_1$ ( pics c \& d) calculated in approximation concerned here.
For $\gamma=5/3$ the corresponding values of $\omega_i$ and $\omega_r$ can be found in fig. 2. 
The decrease of sound speed suppresses an internal gravity modes growth (dashed curves) and its increment vanishes at some 
$\gamma>1$. On the contrary, the surface gravity modes are unstable right up to $\gamma \to 1$. 
Then, as it must be, while $s_1\to 0$  $\omega_i$ of surface mode diminishes up to it's value in homogenerous 
incompressible fluid (Jaroszynski, 1988, SZ1).  Concerning internal mode it's growth disappears at some $s_1 \ne 0$. 

\begin{figure}
\epsfxsize=12cm
\centerline{\epsfbox{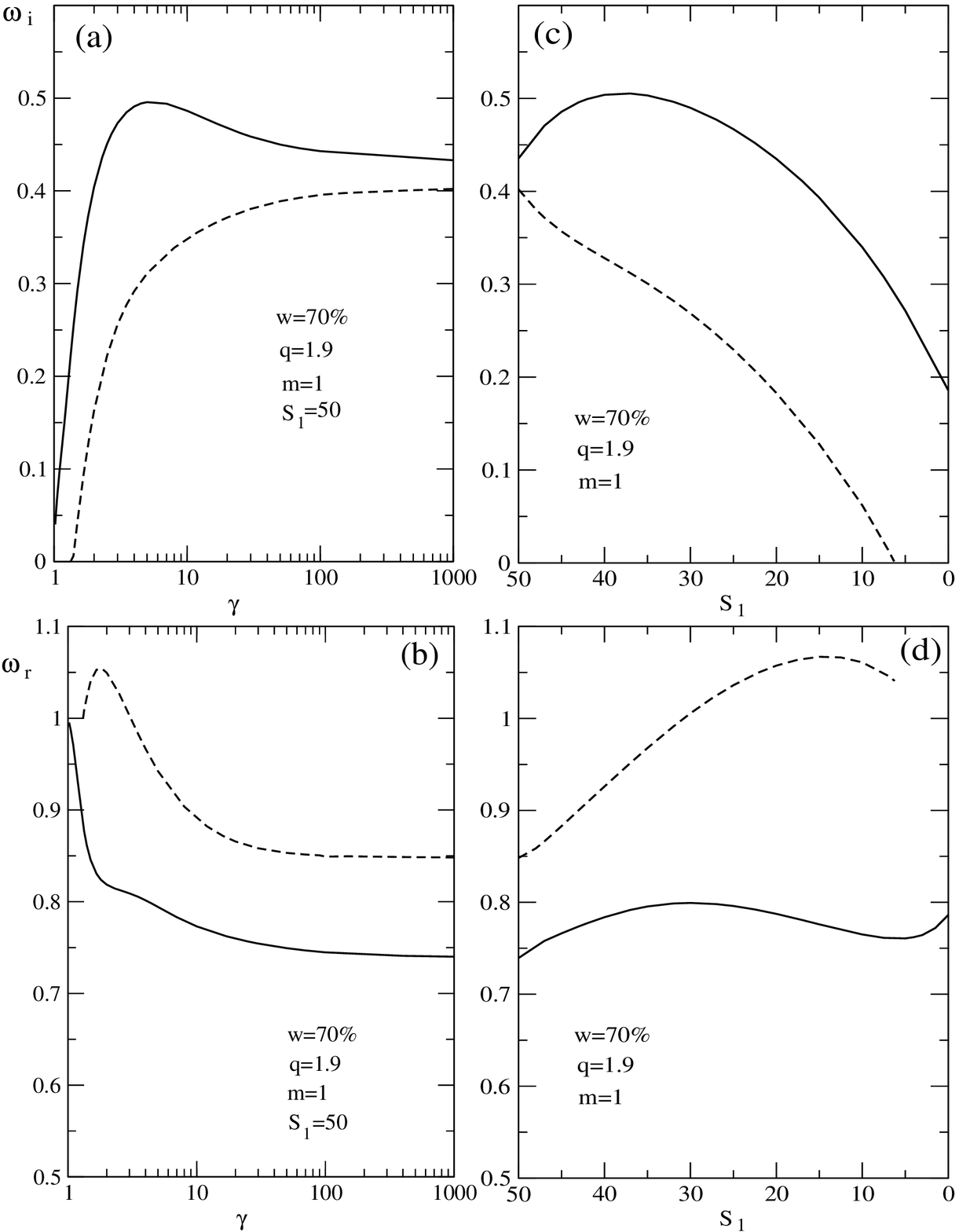}}
\caption
{In pics. (a) and (b) - the dependences $\omega_i(\gamma)$ and $\omega_r(\gamma)$, 
in pics. (c) and (d) - the dependences $\omega_i(s_1)$ and $\omega_r(s_1)$. 
The solid lines denote the growing surface gravity modes, the dashed lines denote the growing 
internal gravity modes.
}
\end{figure}

The parameters in fig.6 correspond to unstable axisymmetric perturbations. We've already noticed that we missed to find 
growing internal gravity modes in the flow with free boundaries and $\gamma<\infty$ that is stable according to Heiland's criterium.
The numerical analysis confirmed this also in incompressible limit. However, it turned out that if someone impose rigid boundaries
the increments of modes that grow with free boundaries increase significantly exceeding  $\Omega_0$.

Moreover, rigid boundaries involve the instability of configuration stable according to Heiland's criterium
(\ref{heiland}). Note that Glatzel (1990) also have found growing internal modes in a problem with rigid boundaries.
At a glance, such influence of boundaries on the unstable internal modes is embarrassing and the stratification must play the main role. 
However one should remember that we consider the amplification of internal oscillations by the shear flow.
The neutral internal modes certainly exist independently on the kind of the boundaries.
But the ability to grow (to decay) for example owing to interaction with the basic flow is controlled not only by a vorticity gradient 
at the critical layer (as it's for homoentropic flow) but additionally by the shape of perturbations field itself as it was shown, in particular, 
by Troitskaya \& Fabrikant (1989) (see also the monograph by Stepanyants \& Fabrikant, 1996). The latter is confirmed also by Glatzel's paper 
(1990), where he found that the internal modes grow (decay) outside the regions of coupling what can happen only if perturbations 
resonantly interact with the basic flow. But he used the simplified angular velocity profile corresponding to the constant 
vorticity in the flow, what excludes the interaction of small perturbations with the flow.

The perturbations profile which is responsible for its growth along with the vorticity gradient in stratified fluid is 
the solution of the boundary problem, so that the amplification of internal modes is determined also by the boundary conditions.

\subsection{ Instability of rotation with quasi-sine deviation from the Keplerian profile}
The instability of flow with such profile was investigated in the previous papers ZS1 and ZS2.
Particularly, it was revealed that the stabilization always 
occurs at some least $K>0$ (look the formula (\ref{rotation11}) ), 
and for $\gamma<\infty$ there's no sonic instability.
The last implies that in non-stratified flow only surface gravity modes can grow.
Accordingly, in the fig.7 we can see how this sort of perturbations is modified by the stratification.
The angular velocity profile is set in order to make equal the maximum enthalpy in the flow with $s_1=0$
to maximum enthalpy in the flow with the same radial size $w$ but rotating with a power law profile  
with $q=1.58$ (see ZS2). 
Note that calculations were carried out for $\gamma=2.5$, since the flow turns out to be stable for ordinary 
$\gamma=5/3$. Several curves $\omega_i(s_1)$ correspond to the different values of $w$. The dependence on $s_1$ is trivial: 
$\omega_i$ increases while the flow becomes less stable according to the Heiland's criterium (\ref{heiland}). 
Similar to the power law rotation the negative entropy gradient ($s_1>0$) widens the range of $w$ when the growing 
surface gravity modes exist.

\begin{figure}
\epsfxsize=12cm
\centerline{\epsfbox{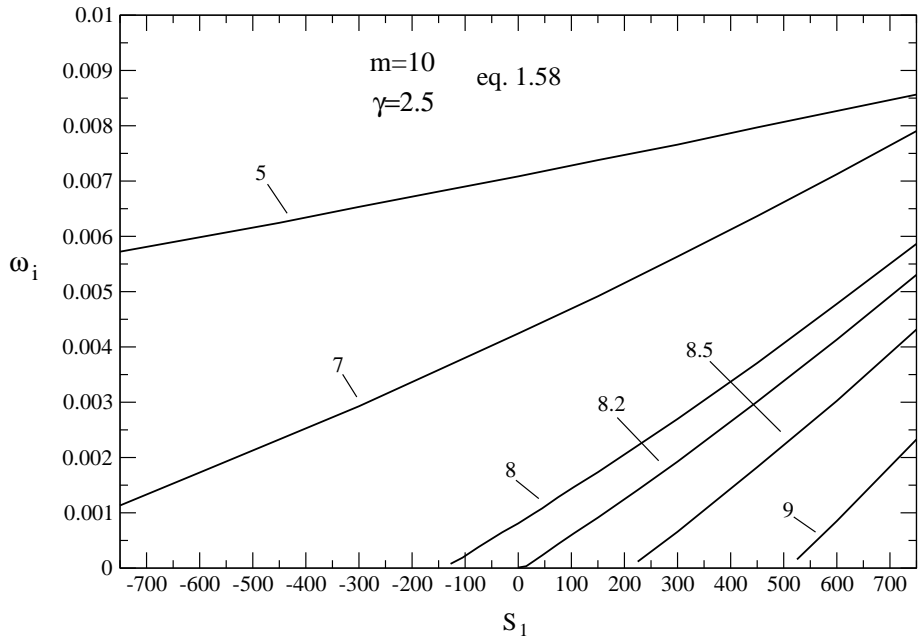}}
\caption
{Increment of the surface gravity mode as a function of 
entropy gradient for the Keplerian rotation with quasi-sine deviation.
The angular velocity profile corresponds to the profile of  
$\Omega(r)$ in the case of power law rotation with the same $w$ (look comments in text). 
In the marked range of $s_1$ the flow is stable to axisymmetric perturbations.
}
\end{figure}

\section{Conclusions}

Well, we investigated in the two-dimensional approximation the instability of the laminar axisymmetric stratified flow with free boundaries to 
infinitesimal non-axisymmetric perturbations. In the case of the power law rotation profile the essential influence of 
entropy gradient on the growing sound and surface gravity modes was revealed. With both negative and positive sign
of entropy gradient relative to ${\bf g}_{eff}$ direction the range of $w$ and $q$ where growing surface gravity modes exist 
becomes wider. Specifically, its increment vanishes at greater $w$ and smaller $q$.   
For sound modes the increments fall into three categories. In the first one $\omega_i$ gradually increases 
while the flow becomes less stable according to the Heiland's criterium (i.e while $s_1$ gets higher). In the second case increment 
vanishes both at some positive ($s_1<0$) and negative ($s_1>0$) entropy gradient. At last, the third category contains 
increments being the result of modes coupling. 
Corresponding $\omega_i(s_1)$ has a break in $s_1^{cr}$, where $\omega_i\to 0$.
Similar peculiarities in increment behavior has been revealed also in ZS2 for 
$\omega_i(w)$ which also had the points $\omega_i\to 0$. 

Besides the sound and surface gravity modes we tried to find growing internal gravity modes. The latter was done 
as well in incompressible limit. It turned out that mentioned sort of perturbations exists only in the flow that 
is unstable according to the Heiland's criterium, i.e. unstable to axisymmetric perturbations. 
However, with rigid boundaries the growing internal modes where found also in the stable to axisymmetric perturbations region (see also Glatzel, 1990).
We suppose such crucial role of the boundaries to be the consequence of specific character of amplification in a stratified medium.
In particular, the transition of energy to perturbations from the basic flow is determined now not only by a vorticity profile in the critical layer but 
also by the profile of perturbations, which in turn depends on the boundaries (Troitskaya \& Fabrikant, 1989). 

In the case of the Keplerian rotation with quasi-sine deviation we found that the increment of the surface gravity mode increases gradually while $s_1$ gets higher. 

We also discussed in the paper that it's incorrect to set the stratification in a problem with free boundaries by a polytropic law with the index different from 
the adiabatic value, since in this case the perturbations don't satisfy the boundary conditions. The last is the consequence of infinite entropy and it's 
derivatives in the boundary points. It reveals that the condition of perturbations regularity at the boundaries is equivalent to the boundary 
condition (i.e. $\Delta p=0$) only if entropy and it's derivatives have finite values at the boundaries. 

\vspace{0.2cm}

\noindent
This paper was supported by grant RFFI-NNIO 06-02-16025.

\end{document}